# Maximum entropy generation rate density and its application to microstructural evolution

Yaw Delali Bensah[1]
March 15, 2017

**Abstract**

A solidification model based on the principle of maximum entropy production rate (MEPR) is considered for the study of pure metals. The approach leads to the development of a breakdown criterion which is able to account for the solidification velocity and solid-liquid interface (SLI) thickness. The quantitative knowledge of the SLI thickness and the maximum entropy generation rate density obtained at breakdown gives an insight about the structure of the SLI during solid to liquid phase transformation. The formation of facet and non-facet morphology, and their transitions are accounted for, which is a function of solidification velocity, heat of fusion, density and the crystallographic growth plane.

## Contents

[1] Department of Materials Science and Engineering, School of Engineering Sciences, University of Ghana, PMB, Accra-Ghana Email: ydbensah@ug.edu.gh/bensahyad@gmail.com.

## 1. Introduction

For many years attempts to theoretically describe the morphological transitions at the solid-liquid interface (SLI) during solidification for pure materials has not been forthcoming. The transitional velocity associated with the solid-liquid transformation especially from a planar interface to a cellular interface which is commonly referred to as the critical/breakdown condition (or breakdown equation) has never been predicted by any known theoretical model for pure materials except with the initial progress made by the maximum entropy production rate (MEPR) model [1, 2]. The difficulty may be partly due to the lack of a comprehensive understanding of the nature and structure of the SLI for pure materials, and the way the SLI thickness fluctuates in response to solidification velocity and crystallographic anisotropy at a fixed temperature gradient.

The proposition that the interface between two adjoining phases has zero thickness (i.e., now labelled as a sharp interface) has been considered by Young [3]; Laplace [4]; Gauss [5]; Wilson [6]; Frenkel [7]; Becker and Doring [8]; Burton, Cabrera and Frank [9]; and others [10]. On the other hand, the consideration of interface diffuseness (i.e., interface of a finite size) during transformation between two phases has been proposed by Rayleigh [11], van der Waals [12], Landau [13] and others [14, 15]. In all, the nature and structure of the existing interface are not well discussed and understood in the latter.

In the early 1930s, Timmermans [16-18] through a series of experiments developed a criterion for distinguishing between plastic crystals and ordered crystals (normal crystals). Though his criterion was not to address issues of SLI during solidification, however, Jackson seemed to have adopted the Timmermans criterion to gain insight into the roughness of an interface on an atomic scale and later provided a theoretical basis for the criterion using the statistical mechanical model of Bragg and Williams. For practical accessibility at the macroscopic scale, the model can predict, the formation of a facet and non-facet morphology for pure materials according to the equation [19]:

$$\alpha_J = \frac{\Delta h_{sl}}{R_g T_m} \tag{1}$$

where $\alpha_J$ (*dimensionless*) is commonly called the Jackson roughness factor, $\Delta h_{sl}$ (*J/mole*) is the heat of fusion, $T_m$ (*K*) is the melting temperature of the material and $R_g$ ($J/moleK$) is the molar gas constant. Jackson deduced that materials with $\alpha_J>2$ will grow to be faceted ($f$) and materials with $\alpha_J<2$ will grow in a non-faceted ($nf$) mode as shown for a number of materials in Table-1. In spite of the success of the Jackson criterion, it is not developed to account for the transition from a facet to non-facet ($f$-$nf$) morphological changes and shows no dependence on the solidification velocity, $V$ ($m/s$), which is one of the most critical parameter during solidification.

Over the years, the terms roughness and diffuseness have both appeared in literature and has been used interchangeably without any clear distinction between them. However, upon examination of the Jackson criterion in equation (1), roughness can be considered to be due to more of thermal influences. Bensah and Sekhar [1, 2] redefined roughness as *thermal diffuseness* given ($\eta_\alpha$) by:

$$\eta_\alpha = \frac{1}{\alpha_J} \qquad (2)$$

Equation (2) is similar to Jackson [20] definition of interface diffuseness.

In an attempt to further understand the nature of the SLI, Cahn [14] introduced a diffuseness parameter $g$ (*dimensionless*) that is to enable the measurement of the number of pseudo atomic layers within the SLI. Cahn projected that when the diffuseness parameter, $g$ which depends on the number of atoms comprising transition from liquid to solid is 1, a sharp interface is obtained and when it changes to become less than 1, then the interface becomes diffuse as given by the expression:

$$g = \frac{\pi^4 \eta_G^3}{8} \exp\left(\frac{-\pi^2 \eta_G}{2}\right) \qquad (3)$$

where $\eta_G$ (*dimensionless*) is the number of pseudo atomic layers (which is the number of lattice planes at the solid-liquid interface). Cahn also gives the expression for the number of pseudo atomic layers at the SLI as:

$$\eta_G = \frac{\zeta}{d} \qquad (4)$$

where $\zeta$ ($m$) is the thickness of the SLI and $d$ ($m$) is the interplanar spacing. Cahn further showed that, the interface diffuseness parameter is a function of the solidification velocity which is the driving force for transformation. However, equations (3) and (4) are difficult to use since the interface thickness is numerically inaccessible either by experiment or by any known theoretical model. From equation (4), the expected dependence of $\zeta$ on the velocity has led to the renaming of $\eta_G$ as the *driving force diffuseness* [1, 2].

In an earlier analyses based on the MEPR model, Bensah and Sekhar [1, 2] have discussed that the formation of facets and non-facets is determined by the value of the *total diffuseness* $\eta_T$ (*dimensionless*) which is a unification of the *driving force diffuseness* and *thermal diffuseness* (Jackson criterion) expressed as:

$$\eta_T = \eta_\alpha + \eta_G \qquad (5)$$

Under the MEPR model [2, 21] Bensah and Sekhar deduced a set of predictive equations for pure materials to account for solid-liquid transition, facet and non-facet formation, and *f-nf* transition, both qualitatively and quantitatively. However, the model was bereft of a breakdown criterion and can only make predictions only when based on experimentally measured breakdown velocity. In this article, we extend the MEPR model and develop a breakdown criterion and equation to make prediction of the solidification velocity and SLI thickness. We particularly consider directional solidification by the Bridgman type solidification technique.

## 2. A thermodynamic basis and theoretical background for MEPR

The principle of MEPR is an extremum approach which states that, if there are sufficient degrees of freedom within a system, it will adopt a stable state at which the entropy generation (production) rate is maximized. Where feasible, the system will also try and adopt a steady state. The MEPR postulate determines the most probable state and therefore allows pathway selections to occur in an open thermodynamic system [15]. While the MEPR postulate was first proposed independently by Ziman [22] and Ziegler [23, 24], we adopt some of the general derivative techniques used by Ziegler to treat liquid-solid transformation during solidification. In this treatment, we consider the SLI to be of finite thickness, $\zeta$ (*m*) which is moving at a velocity $V$ (*m/s*) against an established temperature gradient, $G_{SLI}$ (*K/m*). For the purpose of simplicity, it is assumed that the established temperature gradient, $G_{SLI}$ is at steady state conditions and linear across the SLI according to the relation:

$$\Delta T_{SLI} = G_{SLI} \cdot \zeta \tag{6a}$$

$$\Delta T_{SLI} = T_{li} - T_{si} \tag{6b}$$

where $\Delta T_{SLI}$ is the temperature difference between $T_{li}$ (*K*) and $T_{si}$ (*K*), which are the liquidus and solidus temperatures at the SLI respectively. Since the solidification process is under the influence of a driving force, the Helmholtz free energy per unit volume, $F_H$ (*J/m³*) of the SLI can be written as:

$$F_H = U - T_{av} \, s \tag{7}$$

where $U$ (*J/m³*) is the internal energy per unit volume of the SLI, $s$ (*J/m³K*) is the total entropy per unit volume at the SLI and $T_{av}$ (*K*) is the average temperature at the SLI between $T_{li}$ (*K*) and $T_{si}$ (*K*). For directional solidification, the free energy can be written to be a function of $V$ and $G_{SLI}$ or the cooling rate $\dot{T}$ (K/s) which is given by:

$$\dot{T} = G_{SLI} \, V \tag{8}$$

From equation (8), $\dot{T}$ becomes dependent on $\Delta T_{SLI}$, $G_{SLI}$ and $\zeta$. The velocity and the cooling rate are described as independent state variables, and the time dependent, $\dot{U}(V, \, \dot{T})$, $\dot{s}(V, \, \dot{T})$ and the free energy can be rewritten as:

$$\dot{F}_H(V, \, \dot{T}) = \dot{U} - \dot{T}_{av} \, \dot{s} \tag{9}$$

If the SLI is moving at a force $F$ ($N/m^3$) per unit volume then the total power density $P$ ($J/m^3 s$) transferred is given by:

$$P = F V = \dot{W}_p \tag{10}$$

where $\dot{W}_P$ ($J/m^3 s$) is the work potential rate density stored in the SLI. A combination of the first law of thermodynamics and equation (10) is expressed as:

$$\dot{U} = \dot{Q} + F V = \dot{Q} + \dot{W}_P \tag{11}$$

where $\dot{Q}$ ($J/m^3 s$) is the heat rate density transferred through the SLI. From the second law of thermodynamics, the entropy can be expressed as:

$$\dot{s} T_{av} = \dot{Q} + T_{av} \dot{s}_{gen} \geq 0 \tag{12}$$

The heat rate per unit volume transferred through the SLI is eliminated by combining equations (11) and (12) which gives:

$$\dot{s} T_{av} = \dot{U} - \dot{W}_P + T_{av} \dot{s}_{gen} \geq 0 \tag{13}$$

The total work potential rate density $\dot{W}_P$ ($J/m^3 s$) of the moving SLI can be express as the sum of the work done rate density $\dot{W}_D$ ($J/m^3 s$) and the lost work potential rate density $\dot{W}_L$ ($J/m^3 s$), which is given as:

$$\dot{W}_P = \dot{W}_D + \dot{W}_L \tag{14}$$

Combining equations (13) and (14) gives:

$$\dot{W}_D + \dot{W}_L = \dot{U} - \dot{s} T_{av} + T_{av} \dot{s}_{gen} \geq 0 \tag{15}$$

Equation (15) can be treated by separating the useful work done by the interface $W_D$ (path independent conservative work) from the lost work as:

$$\dot{W}_D = \dot{U} - \dot{s} T_{av} \tag{16}$$

Here we treat the Helmholtz free energy $\dot{F}_H$($J/m^3 s$) as approximately equal to the work done $\dot{W}_D$($J/m^3 s$) and the equation (16) obtained is similar to equation (9). Substitution of equation (16) into equation (15) gives the lost work as:

$$\boldsymbol{\dot{W}_L = T_{av} \, \dot{s}_{gen} \geq 0} \tag{17}$$

The lost work $W_L$ ($J$) is considered as a measure of irreversibility or the degradation of energy from more useful to less useful form. When the lost work reaches a maximum, the work done returns to a minimum, and the work potential become approximately equal to the work lost. The lost work at the SLI is also considered as the amount of work that is irreversibly converted to heat and other related forms. This is also related to the entropy generation across the interface which enables us to express equation (17) as:

$$\boldsymbol{\dot{\varphi}_{max} = \, \dot{s}_{gen} \geq 0} \tag{18}$$

where the expression $\dot{\varphi}_{max}$ ($J/m^3 Ks$) is referred to here as the maximum entropy production rate density (MEPR). The expression $\dot{\varphi}_{max}$ could possibly have a link to the dissipative function $\Phi$ which was first introduced by Raleigh [25] and later used by Onsager [26], Prigogine [27, 28] and Ziegler [29, 30]. However, the connection between the maximum entropy production rate density and the dissipative function is left for future study.

### 3. Model and entropy balance across a solid-liquid interface

Considering a one dimensional treatment for the entropy balance across the SLI at steady state conditions, where the maximum entropy production rate density is given by [1, 2]:

$$\dot{\varphi}_{max} = \dot{s}_E - \dot{s}_{LG} \tag{19}$$

where $\dot{s}_E$ $(J/m^3Ks)$ is the change in entropy generation rate density which describes the new entropy generated due to exchange of matter and energy to and from the SLI with the surrounding as expressed in equation (21) [1, 2, 15, 31] and $\dot{s}_{LG}$ $(J/m^3Ks)$ is the entropy generation rate density which describes the force-flux entropy generated by the solute gradient in the liquid as expressed in equation (22) [1, 2, 15, 31]. The maximum entropy generation rate density (MEPR) is achieved when the moving interface losses work due to entropy generation through heat dissipation which is given [1, 2]:

$$\dot{\varphi}_{max} = \frac{\Delta\rho_k}{2} \frac{V^3}{\zeta^2 G_{SLI}} \tag{20}$$

where $\Delta\rho_k$ $(kg/m^3)$ is the overall density shrinkage associated with liquid to solid transformation expressed as $\Delta\rho_k = \rho_l \, \Delta\rho \, / \rho_s$, and $\Delta\rho$ $(kg/m^3)$ is the density change from liquid to solid $(\rho_s - \rho_l)$; $\rho_s$ $(kg/m^3)$ and $\rho_l$ $(kg/m^3)$ are the densities of rigorous solid and liquid respectively. The $\dot{s}_E$ $(J/m^3Ks)$ is given as [1, 2, 15, 31]:

$$\dot{s}_E = \frac{V \, \Delta h_{sl} \, G_{SLI}}{T_{li} \cdot T_{si}} \tag{21}$$

where $\Delta h_{sl}$ $(J/m^3)$ is the equilibrium heat of fusion. The $\dot{s}_{LG}$ $(J/m^3Ks)$ is given as [1, 2, 15, 31]:

$$\dot{s}_{LG} = \frac{V^2 C_O \, R_g}{4 \, D_L} \frac{\ln(1/k_{eff}) \, (1 - k_{eff})}{k_{eff}} \tag{22}$$

where $C_O$ $(mole/m^3)$ is the initial solute concentration in the liquid, $R_g$ $(J/mole \, K)$ is the gas constant, $D_L$ $(m^2/s)$ is the coefficient of diffusion of solute and $k_{eff}$ $(dimensionless)$ is the effective partition coefficient. For pure materials solute partitioning is absent and equation (22) becomes zero when the expression $(\ln(1/k_{eff}) \, (1 - k_{eff})/k_{eff})$ turns zero as $k_{eff}$ becomes equal to one. Combining equations (19) to (21) gives the SLI thickness for a pure material as:

$$\zeta = \frac{V}{G_{SLI}} \left( \frac{\Delta\rho_k \, T_{si} \, T_{li}}{2 \, \Delta h_{sl}} \right)^{1/2} \tag{23}$$

The parameters $T_{si}$ and $T_{li}$ in equation (23) are not readily known but can be approximated as $T_{si}=T_m$ and $T_{li}=T_m$. Equation (23), then becomes:

$$\zeta = \frac{V}{G_{SLI}} \left( \frac{\Delta\rho_k \, T_m^2}{2 \, \Delta h_{sl}} \right)^{1/2} \tag{24}$$

The temperature gradient at the SLI has been approximately defined as [1, 2, 15, 31]:

$$G_{SLI} = \frac{G_L + G_S}{2} \tag{25}$$

In directional solidification it is conventional that the imposed temperature gradient is across the liquid melt. If the microstructural growth at the SLI is into the liquid and the heat flow into the fully formed solid are opposite, then it becomes logical to propose that the temperature gradient of the liquid melt and that of the rigorous liquid component within the SLI be treated to be approximately equal. Furthermore, the flow of heat into the rigorous solid within the SLI is expected to be distributed along the *hkl* of the chosen crystallographic plane for which growth takes place. It becomes convenient and simplistic to therefore define the temperature gradient of the rigorous solid within the SLI as:

$$G_S = G_L \theta \qquad (26)$$

where $\theta$ (*dimensionless*) is the chosen crystallographic plane ($h^2+k^2+l^2$) of growth by the rigorous solid within the SLI. Putting equations (25) and (26) into equation (24) gives:

$$\zeta = \frac{2V}{G_L(1+\theta)} \left( \frac{\Delta \rho_k}{2} \frac{T_m^2}{\Delta h_{sl}} \right)^{1/2} \qquad (27)$$

It should be noted that, no breakdown criterion has been established based on equations (20) and (21) and therefore, are informative only before and after breakdown, unless otherwise the behaviour/expression of $\zeta$ and $V$ are known at breakdown and beyond. Likewise, the expected linear relationship between $\zeta$ and $V$ in equation (27) is useful at all solidification conditions i.e., before breakdown and beyond.

## 4. Mixing entropies in liquid melt and solid-liquid interface

We consider the entropy associated with the self-diffusion of the atomic particles in the liquid melt which can be governed by the well-known Einstein-Stokes equation. Also within the interface, entropy is generated through the mixing of the rigorous liquid and rigorous solid which can be analysed by solution thermodynamics.

Considering the boundary between the fully liquid melt zone and the SLI, the flow of particles from the liquid melt into the SLI and finally to the fully formed solid is accompanied by a net entropy change which can be given by:

$$\dot{\sigma}_{net} = \dot{s}_K - \dot{s}_{ES} \qquad (28)$$

where $\dot{\sigma}_{net}$ ($J/m^3Ks$) is the net mixing entropy rate density, $\dot{s}_K$ ($J/m^3Ks$) is the Sekhar entropy rate density which is the entropy associated with the mixing of a rigorous solid and rigorous liquid in the SLI and $\dot{s}_{ES}$ ($J/m^3Ks$) is the Einstein-Stokes entropy rate density which is the entropy associated with the viscous flow of a liquid melt into the SLI. It should be noted that though the Einstein-Stokes entropy rate density is diffusional entropy it can also be considered as mixing entropy due to the spatial distribution of the liquid melt particles as a result of the associated diffusion gradient. Thus, the Sekhar entropy rate density is analogous to the Einstein-Stokes entropy rate density.

From equation (28), we assume that the velocities of the atomic particles in the liquid melt close to the SLI and that of the moving SLI are approximately equal. From equation (28), we establish that the net mixing entropy rate density is equal to zero for a derivative of velocity at the peak if considered a parabolic curve.

## 4.1 Sekhar entropy rate density

Now considering the SLI, let $f_s$ (*dimensionless*) and $f_l$ (*dimensionless*) be the fractions of the rigorous solid and rigorous liquid expressed as:

$$f_s + f_l = 1 \tag{29a}$$

The change in the fractions of the rigorous solid and rigorous liquid at the SLI is also expressed as:

$$df_s + df_l = 1 \tag{29b}$$

On the basis of the solidification velocity and SLI thickness, the change in fraction solidified with respect to change in time is given as [1, 2, 15, 31]:

$$\frac{df_s}{dt} = \frac{V}{\zeta} \tag{30}$$

The change in the entropy of mixing $dS_{mix}$ (*J/K*) of the rigorous solid fraction and rigorous liquid fraction at the SLI can be written as:

$$dS_{mix} = -R_g[f_s \ln f_s + f_l \ln f_l] \, dn_s \tag{31}$$

where $n_s$ (*mole*) is the number of moles of the rigorous solid at the SLI. Multiplying equation (31) by equations (30) and (29b) gives:

$$\frac{df_s \, (1 - df_l) \, dS_{mix}}{dn \, dt} = -\frac{VR_g}{\zeta}\left[f_s \ln f_s + f_l \ln f_l\right] df_s \tag{32}$$

Letting ($dS_{mix}/dt \, dn$) be $\dot{S}_{fs}$ (*J/mole K s*) and integrating equation (32) gives:

$$\dot{S}_{fs} \int_0^{f_s} df_s - df_s \, df_l = -\frac{VR_g}{\zeta} \int_0^1 \left(f_s \ln f_s + f_l \ln f_l\right) df_s \tag{33}$$

$$\dot{S}_{fs} = \frac{VR_g}{4\zeta} \tag{34}$$

For any given pure material, the change in the mixing entropy generation rate density $\dot{s}_{fs}$ (*J/m³Ks*) of the rigorous solid at the SLI is given as:

$$\dot{s}_{fs} = \frac{V \, R_g \, \rho_s}{4 \, \zeta A_w} \tag{35}$$

where $A_w$ (*Kg/mole*) is the atomic weight of the pure material. Note that the result for equation (35) was first obtained in a seminal paper by Sekhar [15] but in a sketchy manner which is twice the results obtained in equation (35) due to certain approximations used in his approach. If the procedure from equations (29-34) is repeated for the fraction of the rigorous liquid ($f_l$) at the SLI then, the change in the mixing entropy generation rate density $\dot{s}_{fl}$ (*J/m³Ks*) of the rigorous liquid at the SLI is obtained as:

$$\dot{s}_{fl} = \frac{V \, R_g \, \rho_l}{4 \, \zeta A_w} \tag{36}$$

The sum of equations (35) and (36) is total change in entropy rate density for both the rigorous liquid and the rigorous solid at the SLI and is expressed as:

$$\dot{s}_K = \frac{V \, R_g \, (\rho_l + \rho_s)}{4 \, \zeta A_w} \tag{37}$$

## 4.2 Einstein-Stokes entropy rate density

Considering again, a molten pure metal in which its spherical particles (atoms/molecules) are in motion in its own fluid and are non-reacting. At a steady state condition, the shear viscosity, $S_\eta$ ($Js/m^3$) of the fluid is given as:

$$S_\eta = \frac{R_g T_m}{6\pi D_S r} \tag{38}$$

where $D_S$ ($m^2/s$) is the coefficient of diffusion of the atomic particles of a pure metal and $r$ ($m$) is the radius of a spherical atomic particle. In the melt, the atomic (or molecular) acceleration per unit temperature $a_g$ ($mole/s^2 K$) against the viscous liquid melt can be expressed as:

$$a_g = \left(\frac{\rho_l}{A_w}\right)\left(\frac{V^2}{G_L}\right) \tag{39}$$

The first term in parenthesis of equation (39) is the equivalent molar concentration ($mole/m^3$) and the second term is the volumetric acceleration per unit temperature ($m^3/s^2 K$) of the atomic particles of the pure material. Equation (38) can be transformed to entropy generation rate density generated due to viscous flow, $\dot{s}_{ES}$ ($J/m^3 Ks$) of the particles in the fluid by multiplying with equation (39) to give:

$$\dot{s}_{ES} = \frac{R_g T_m \rho_l V^2}{6\pi D_S r A_w G_L} \tag{40}$$

## 5. Interface breakdown criterion

Based on equation (28), we can establish that interface breakdown occurs according to the equation:

$$\left(\frac{\partial \dot{\sigma}_{net}}{\partial V}\right)_\zeta = 0 \tag{41}$$

Putting equations (37) and (40) into equation (28), and applying the breakdown criterion in equation (41) gives:

$$\left(\frac{\partial \dot{\sigma}_{net}}{\partial V}\right)_\zeta = \frac{R_g\,(\rho_l + \rho_s)}{4\,\zeta A_w} - \frac{V R_g T_m \rho_l}{3\pi D_S r A_w G_L} = 0 \tag{42}$$

From equation (42) the SLI thickness at breakdown is obtained as:

$$\zeta_C = \frac{3\pi D_S r\,(\rho_l + \rho_s)\,G_L}{4\,T_m V_C \rho_l} \tag{43}$$

From equation (43), the SLI thickness is still a function of the velocity i.e. equation (43) has two unknown parameters just as equation (27). Having defined the scope of application for equation (27), and combining it with equation (43) gives an approximate point of intersection that produces breakdown expression for the SLI thickness independent of the velocity as:

$$\zeta_C = \left(\frac{9}{8}\right)^{1/4}\frac{(\pi D_S r)^{1/2}\,(\rho_l + \rho_s)^{1/2}\,\Delta\rho_k^{1/4}}{(\Delta h_{sl})^{1/4}\,\rho_l^{1/2}\,(1+\theta_C)^{1/2}} \tag{44}$$

where $\theta_C$ is equal to $(1^2+1^2+1^2)$ for FCC materials, $(1^2+1^2+0^2)$ for BCC materials, etc., when closed packed planes are considered.

In the case of pure FCC material the SLI thickness at breakdown is given as:

$$\zeta_C = 0.515 \frac{(\pi D_S r)^{1/2} (\rho_l+\rho_s)^{1/2} \Delta \rho_k^{1/4}}{(\Delta h_{sl})^{1/4} \rho_l^{1/2}} \tag{45}$$

And for pure BCC materials the SLI thickness at breakdown is given as:

$$\zeta_C = 0.595 \frac{(\pi D_S r)^{1/2} (\rho_l+\rho_s)^{1/2} \Delta \rho_k^{1/4}}{(\Delta h_{sl})^{1/4} \rho_l^{1/2}} \tag{46}$$

From the same equations (27) and (43), one is able to obtain the breakdown solidification velocity as:

$$V_C = \left(\frac{9}{32}\right)^{1/4} \cdot \frac{1}{T_m} \cdot G_L (\pi D_S r)^{1/2} (\Delta h_{sl})^{1/4} \left[\frac{(\rho_l+\rho_s)^{1/2}}{\rho_l^{1/2} \Delta \rho_k^{1/4}}\right] (1+\theta_C)^{1/2} \tag{47}$$

For pure FCC materials the breakdown velocity is given as:

$$V_C = \frac{1.456}{T_m} \cdot G_L (\pi D_S r)^{1/2} (\Delta h_{sl})^{1/4} \left[\frac{(\rho_l+\rho_s)^{1/2}}{\rho_l^{1/2} \Delta \rho_k^{1/4}}\right] \tag{48}$$

And for pure BCC materials the breakdown velocity is given as:

$$V_C = \frac{1.261}{T_m} \cdot G_L (\pi D_S r)^{1/2} (\Delta h_{sl})^{1/4} \left[\frac{(\rho_l+\rho_s)^{1/2}}{\rho_l^{1/2} \Delta \rho_k^{1/4}}\right] \tag{49}$$

## 6. Maximum entropy generation rate density and total diffuseness at breakdown

Up to this point we have been able to obtain breakdown criterion and equation from the knowledge of the Sekhar entropy rate density and the Einstein-Stokes entropy rate density. For the case of pure materials, the maximum entropy generation rate density has a common expression given as:

$$\dot{\varphi}_{max} = \dot{s}_E = \left(\frac{\Delta h_{sl} G_{SLI}}{T_m^2}\right) V \tag{50}$$

Equation (50) is linear with changing values of the solidification velocity and is valid at all velocities, that is, when $0<V>V_C$. Though equation (43) has been derived for breakdown condition, it is only a certain value of the velocity which can lead to interface breakdown when it is varied. Once breakdown is established, the maximum entropy generation rate density can be obtained by substituting equation (43) into equation (20) to give:

$$\dot{\varphi}_{max} = \left(\frac{16 \Delta \rho_k T_m^2 \rho_l^2}{9 (\pi D_S r)^2 G_L^3 (\rho_l+\rho_s)^2 (1+\theta)}\right) V^5 \tag{51}$$

When equations (50) and (51) are plotted against the solidification velocity, the crossover point between the two equations is connected and approximately equal to the breakdown of the SLI. The crossover velocity would be the same as equation (47). Further at the crossover, the maximum entropy generation rate density obtained, is equivalent to the breakdown equation given in equation (52), which is obtained by putting equations (44) and (47) into equation (20).

$$(\dot{\varphi}_{max})_C = 0.364 \; \frac{G_L^2 \, \Delta h_{sl}^{5/4} \, (\pi D_S r)^{1/2} \, (\rho_l + \rho_s)^2 \, (1+\theta)^{3/2}}{\Delta \rho_k^{1/4} \, T_m^3 \, \rho_l^2} \tag{52}$$

For any given interface thickness the *driving force diffuseness*, $\eta_G$ (*dimensionless*) which describes the number of pseudo atomic layers within the SLI region given in equation (4) [1, 2, 14, 15, 31] combined with equation (27), and with equation (43), gives the number of pseudo atomic layers at the SLI before and after breakdown respectively as:

$$\eta_G = \frac{2V}{G_L(1+\theta)} \left( \frac{\Delta \rho \, T_m^2}{2 \, \Delta h_{sl} \, d^2} \right)^{1/2} \tag{53}$$

$$(\eta_G)_C = \left( \frac{9}{8} \right)^{1/4} \frac{(\pi D_S r)^{1/2} \, (\rho_l + \rho_s)^{1/2} \, \Delta \rho_k^{1/4}}{(\Delta h_{sl})^{1/4} \, \rho_l^{1/2} \, (1+\theta_C)^{1/2} \, d} \tag{54}$$

Similarly, combining equations (2), (5) and (53), and, equations (2), (5) and (54), gives the *total diffuseness* before and at breakdown respectively as:

$$\eta_T = \frac{2V}{G_L(1+\theta)} \left( \frac{\Delta \rho \, T_m^2}{2 \, \Delta h_{sl} \, d^2} \right)^{1/2} + \frac{\Delta h_{sl}}{R_g \, T_m} \tag{55}$$

$$(\eta_T)_C = \left( \frac{9}{8} \right)^{1/4} \frac{(\pi D_S r)^{1/2} \, (\rho_l + \rho_s)^{1/2} \, \Delta \rho_k^{1/4}}{(\Delta h_{sl})^{1/4} \, \rho_l^{1/2} \, (1+\theta_C)^{1/2} \, d} + \frac{\Delta h_{sl}}{R_g \, T_m} \tag{56}$$

## 7. Results and discussion

Under the MEPR approach, a criterion for interface breakdown for pure metals has been established according to equation (41). The breakdown criterion given in equation (41) is applied to equation (28) which is represented graphically at the peak of Figure-2. For a number of pure metals considered for in Table-1, the breakdown velocity calculated from equation (47) are all in the order of a micron per second and is a strong function of the coefficient of diffusion, atomic radius, heat of fusion and the temperature gradient.

The general expression for the SLI thickness given in equation (27) is useful before and after breakdown conditions. At breakdown, the SLI thickness derived is given in equation (44) and their calculated values are shown in Table-1 for a number of pure materials. It is noted that the SLI thickness at breakdown is independent of the temperature gradient. In all the materials given in Table-1, the SLI thickness is less than the atomic radius, the lattice parameter and/or the interplanar spacing. The significance of this result is that, the SLI thickness calculated has all to do with space and not matter; i.e. the SLI is empty and therefore contains no liquid atoms and/or solid crystals. In other words, there is no mixture and/or of rigorous liquid (atoms) or rigorous solid (crystals) at the SLI. The SLI at breakdown is only a function of materials constants and the crystallographic growth plane chosen by the interface and as such it is cannot fundamentally be zero. Furthermore, the calculated values of SLI thickness in Table-1 shows that the density changes across the interface are not necessarily discontinuous and that atoms from the rigorous liquid hop across the interface to the rigorous solid region. It is therefore reasonable to infer that all pure single element materials will have an interface size smaller than the atomic radius and the interplanar spacing.

Under this model, the *thermal diffuseness* and the *driving force diffuseness* are shown to be unified through the *total diffuseness* for the prediction of facets and non-facets formation. For pure materials a non-facet morphology is formed when $\eta_T > 0.5$ whiles a facet morphology is formed when $\eta_T < 0.5$ as seen in Table-1. When $\eta_T = 0.5$, then the material has the tendency to form facet and non-facet morphology. The results are in direct agreement with known and available experimental observations [32-39].

In the MEPR model, the maximum entropy generation rate density (equation 20) is fundamentally important to the study of SLI breakdown and plays a major role in accessing the structure of the SLI for our understanding of solidification. While it is formulated to obey the second law of thermodynamics at all conditions, it is always positive. Negative values are forbidden and it is not expected to approach $\pm\infty$. It can attain a value of zero only at zero solidification velocity. A zero maximum entropy production rate density means it is at a thermodynamic equilibrium. The maximum entropy generation rate density is a measure of atomistic level entropy generated per unit time for SLI for matter (when $\zeta_C > r$ or $\zeta_C > d$) and space (when $\zeta_C < r$). In other words, equation (20) is a dual type of equation that can evaluate space and matter at the SLI. In the results given in Table-1, the maximum entropy production rate density at breakdown measures only space. This result could be akin to the Perelman entropy functional ($W$) that deals with the measurement of disorder in the global geometry of 3-dimensional space which was employed as a tool in the theory of Ricci flow for studying geometric curvature in 3-dimensional manifolds [40]. Though geometrically different and the realms of applications are quite dissimilar, the fundamental concepts could be connected. This comparison is for now based on supposition and is left for future study.

Figure-1 which was obtained from the plot of equations (50) and (51) against the velocity produces a crossover point which is equivalent to the breakdown condition obtained from equation (49). The plot obtained for Figure-1 is also a confirmation of the schematic predictions made by Sekhar [15]. The same graphical results were earlier on obtained by Hill [41] in the study of solidification for $NH_4Cl$ by an application of a different extremum principle.

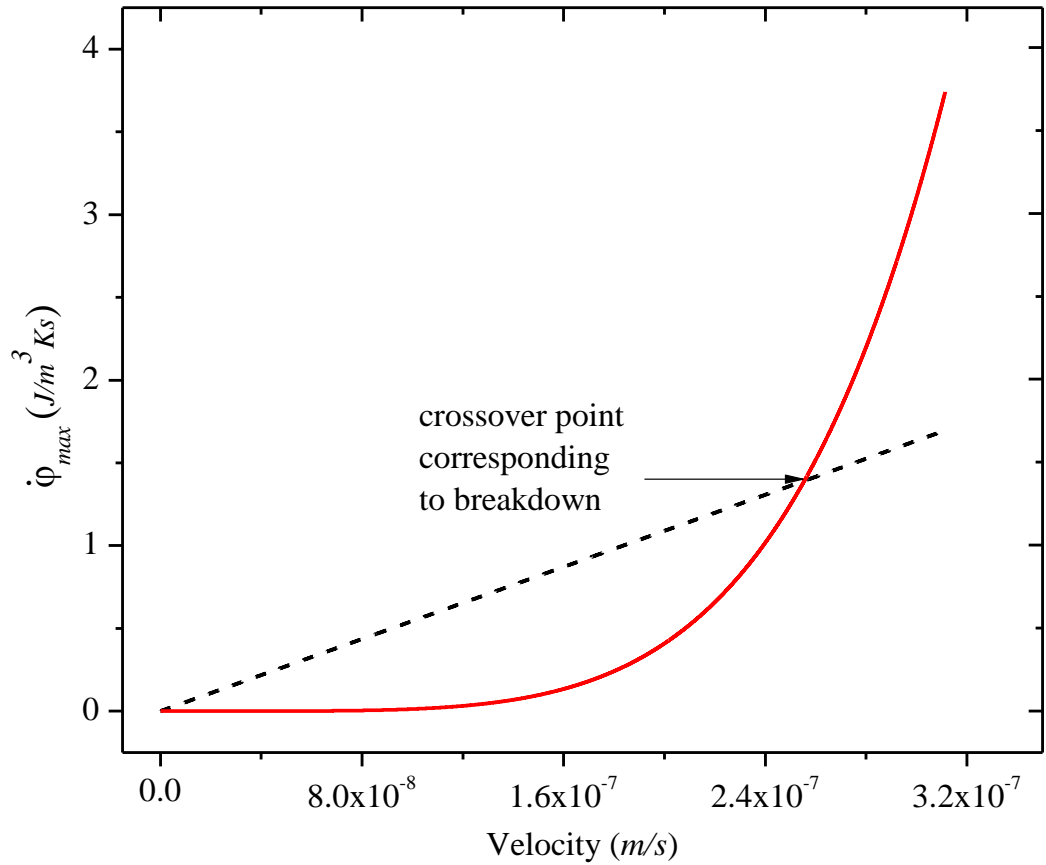

**Figure 1:** A plot showing the maximum entropy production rate density as against the velocity. The dotted black line represent equation (50) and the full red line represent equation (51). The crossover point for the two lines corresponds to the interface breakdown.

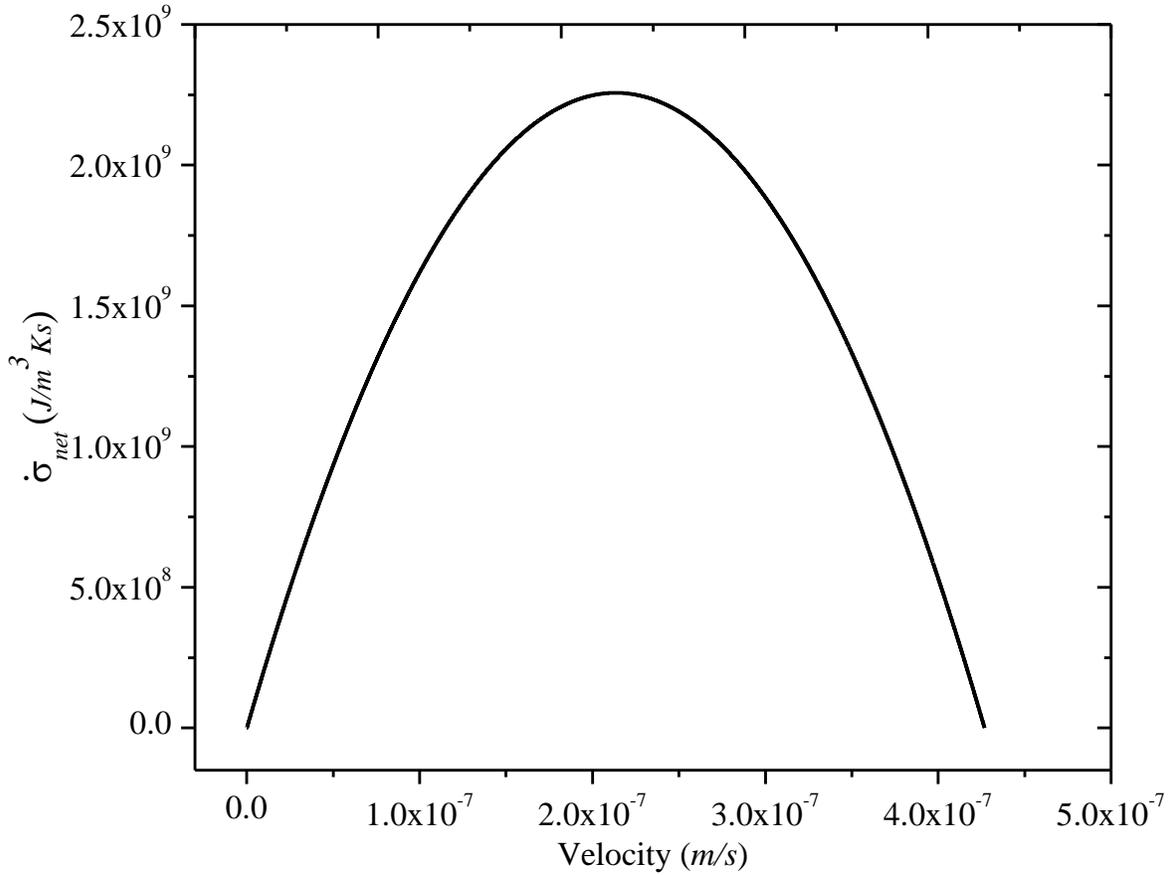

**Figure 2:** A plot of the net entropy production rate density against the velocity of the interface as given in equation (28) given for copper metal. The peak of the curve which corresponds to point where the first derivative is zero represents the breakdown as derived by equation (41). The peak point is the same as the crossover point in figure-1.

**Table-1:** Model calculations for selected pure metals. The temperature gradient value used is 3000 $K/m$.

| Material | Jackson criterion | | Material properties | | MEPR predictions | | | | | |
|---|---|---|---|---|---|---|---|---|---|---|
| | $\alpha_J$ | $f/nf$ prediction | $D(T_m)_S$ $(10^{-9}\,m^2/s)$ | $r$ $(10^{-10}\,m)$ | $V_C$ $(10^{-6}\,m)$ | $\zeta_C$ $(10^{-9}\,m)$ | $(\dot{\varphi}_{max})_C$ $(J/m^3 s)$ | $\eta_a$ (no units) | $\eta_G$ (no units) | $\eta_T$ (no units) |
| Bi | 2.50 | $f$ | 0.80 [42] | 1.56 | 0.17 | 0.019 | 0.896 | 0.401 | 0.039 | 0.440 |
| Pb | 0.96 | $nf$ | 2.19 [43] | 1.75 | 0.33 | 0.028 | 1.399 | 1.047 | 0.097 | 1.144 |
| Tin | 1.67 | $nf$ | 2.05 [42] | 1.40 | 0.25 | 0.029 | 2.426 | 0.597 | 0.078 | 0.676 |
| Ge | 3.67 | $f$ | 12.1 [44] | 1.22 | 0.56 | 0.030 | 6.538 | 0.273 | 0.091 | 0.364 |
| Li | 0.79 | $nf$ | 6.80 [45] | 1.52 | 1.53 | 0.022 | 7.413 | 1.257 | 0.089 | 1.346 |
| Na | 0.84 | $nf$ | 4.19 [46] | 1.86 | 1.08 | 0.028 | 3.703 | 1.186 | 0.094 | 1.279 |
| Rb | 0.84 | $nf$ | 2.62 [47] | 2.48 | 0.71 | 0.042 | 1.230 | 1.186 | 0.106 | 1.292 |
| Ni | 1.22 | $nf$ | 4.60 [48] | 1.24 | 0.22 | 0.022 | 1.029 | 0.822 | 0.109 | 0.931 |
| Cu | 1.17 | $nf$ | 3.97 [49] | 1.28 | 0.26 | 0.021 | 1.392 | 0.851 | 0.103 | 0.954 |
| Ag | 1.09 | $nf$ | 2.56 [50] | 1.44 | 0.19 | 0.022 | 0.753 | 0.910 | 0.094 | 1.005 |
| Cs | 0.83 | $nf$ | 2.69 [51] | 2.65 | 0.68 | 0.049 | 0.979 | 1.200 | 0.173 | 1.373 |
| Ga | 2.22 | $f$ | 1.60 [51] | 1.35 | 0.47 | 0.021 | 7.439 | 0.451 | 0.052 | 0.503 |
| In | 0.92 | $nf$ | 1.68 [51] | 1.67 | 0.31 | 0.029 | 1.022 | 1.089 | 0.073 | 1.162 |
| Tl | 0.86 | $nf$ | 2.01 [51] | 1.70 | 0.23 | 0.037 | 0.475 | 1.159 | 0.089 | 1.249 |
| Zn | 1.27 | $nf$ | 2.03 [51] | 1.34 | 0.23 | 0.024 | 1.079 | 0.787 | 0.059 | 0.846 |

## 8. Conclusion

By MEPR model we have been able to arrive at a breakdown equation that enables the prediction of the SLI breakdown solidification velocity and the SLI thickness for pure materials. The SLI thickness is of finite size and cannot be zero. The SLI can be described as a diffuse interface as far as there is a finite gap even if devoid of rigorous liquid and/or solid. Generally, all pure metallic materials (elements) are pointing to the direction of SLI thickness less than both the atomic size and the interplanar spacing. It is now conceptually clear that the maximum entropy generation rate density can evaluate and give a qualitative view of the structure of the SLI.

**Acknowledgement**

This work was supported through personal funds. Thanks to Dr. J.A. Sekhar (professor emeritus at the University of Cincinnati, Cincinnati, Ohio, USA) who first introduced me to this field of study.